\documentclass[conference]{IEEEtran}
\IEEEoverridecommandlockouts

\usepackage{amsmath,amssymb,amsfonts}
\usepackage{algorithmic}
\usepackage{graphicx}
\usepackage{textcomp}
\usepackage{multirow}
\usepackage{booktabs}
\usepackage{makecell}
\usepackage{colortbl}
\usepackage{cite}
\usepackage{bm}
\usepackage{pifont}
\usepackage[dvipsnames]{xcolor} 
\usepackage[colorlinks=true,  linkcolor=black, urlcolor=black, citecolor=black, urlcolor=Magenta]{hyperref} 

\def\BibTeX{{\rm B\kern-.05em{\sc i\kern-.025em b}\kern-.08em
    T\kern-.1667em\lower.7ex\hbox{E}\kern-.125emX}}
\begin{document}

\title{Spectral-Aware Low-Rank Adaptation for Speaker Verification

\thanks{Thanks to Research Grants Council of Hong Kong, Theme-based Research Scheme (Ref.: T45-407/19-N).}
}

\author{
    \IEEEauthorblockN{Zhe Li}
    \IEEEauthorblockA{
        \textit{Dept. of Electrical and Electronic Engineering} \\
        \textit{The Hong Kong Polytechnic University}\\
        lizhe.li@connect.polyu.hk
    }
    \and
    \IEEEauthorblockN{Man-wai Mak}
    \IEEEauthorblockA{
        \textit{Dept. of Electrical and Electronic Engineering} \\
        \textit{The Hong Kong Polytechnic University}\\
        enmwmak@polyu.edu.hk
    }
    \and
    \IEEEauthorblockN{Mert Pilanci}
    \IEEEauthorblockA{
        \textit{Dept. of Electrical Engineering} \\
        \textit{Stanford University}\\
        pilanci@stanford.edu
    }
    \and
    \IEEEauthorblockN{Hung-yi Lee}
    \IEEEauthorblockA{\textit{Dept. of Electrical Engineering} \\
\textit{National Taiwan University}\\
hungyilee@ntu.edu.tw}
\and
\IEEEauthorblockN{Helen Meng}
\IEEEauthorblockA{\textit{Dept. of Systems Engineering \& Engineering Management} \\
\textit{The Chinese University of Hong Kong} \\
hmmeng@se.cuhk.edu.hk}
}

\maketitle

\begin{abstract}
Previous research has shown that the principal singular vectors of a pre-trained model's weight matrices capture critical knowledge. In contrast, those associated with small singular values may contain noise or less reliable information. As a result, the LoRA-based parameter-efficient fine-tuning (PEFT) approach, which does not constrain the use of the spectral space, may not be effective for tasks that demand high representation capacity. In this study, we enhance existing PEFT techniques by incorporating the spectral information of pre-trained weight matrices into the fine-tuning process. We investigate spectral adaptation strategies with a particular focus on the additive adjustment of top singular vectors. This is accomplished by applying singular value decomposition (SVD) to the pre-trained weight matrices and restricting the fine-tuning within the top spectral space. Extensive speaker verification experiments on VoxCeleb1 and CN-Celeb1 demonstrate enhanced tuning performance with the proposed approach. Code is released at \url{ https://github.com/lizhepolyu/SpectralFT}.
\end{abstract}


\section{Introduction}
The primary goal of parameter-efficient fine-tuning (PEFT) is to reduce the number of tunable parameters compared to full fine-tuning. This approach conserves computational resources and enables easy sharing of lightweight, fine-tuned models \cite{peng2023parameter,sang2024efficient,li2024dual}. Among these methods, the low-rank adaptation (LoRA) model \cite{hulora} stands out for its simplicity and effectiveness. LoRA tunes an additional, trainable low-rank matrix, resulting in zero inference latency after integrating the adapter into the pre-trained model. Since its introduction, several LoRA variants have emerged. For instance, AdaLoRA \cite{zhang2023adaptive}, IncreLoRA \cite{zhang2023increlora}, and DyLoRA \cite{valipour2023dylora} dynamically adjust the rank of the LoRA adaptation matrices to enhance tuning efficiency. A more recent variant, DoRA \cite{liudora}, decomposes a pre-trained weight matrix into a magnitude vector and a series of direction vectors.

Although LoRA is simple and effective, its low-rank constraint may be suboptimal for tasks that demand high representation capacity. In particular, for a rank $r$ approximation of a matrix $\bm{W}$, the optimal solution corresponds to the largest $r$ singular values and their corresponding singular vectors—components that LoRA does not explicitly leverage. This limitation implies that potentially valuable directions in the parameter space, captured by these singular vectors, remain underutilized.

Previous research, such as \cite{zhang2024spectral,gao2024adaptive,meng2024pissa,wang2024milora}, explored incorporating the spectral information from the pre-trained model's weight matrices into PEFT by introducing a spectral adaptation mechanism that updates the top singular vectors of the pre-trained weight matrices. Other studies \cite{zhang2024spectrum,lilorap,yang2024corda,nikdanrosa,hameed2024rosa} further exploited the spectral space of pre-trained weight matrices, adjusting both singular values and singular vectors during fine-tuning. These approaches focus on the spectral components' magnitude and directions, aiming for a more refined and effective adaptation. Collectively, these works contribute to a deeper understanding of the relationship between the spectral information of weight matrices and model performance. In this work, we leverage the spectral information of the pre-trained weight matrices during fine-tuning to enhance the model’s performance.

This paper introduces a spectral fine-tuning (SpectralFT) method based on low-rank adaptation to adapt a pre-trained Transformer-based speech model for speaker verification. Specifically, we decompose a weight matrix $\bm{W}$ using singular value decomposition (SVD). Based on the magnitude of the singular values, $\bm{W}$ is divided into two components: a principal matrix $\bm{W}_p$, associated with the larger singular values, and a minor matrix $\bm{W}_m$, associated with the smaller singular values. The principal matrix encapsulates the core of the pre-trained knowledge, and we approximate the original parameter matrix $\bm{W}$ using this low-rank matrix $\bm{W}_p$. The principal matrix $\bm{W}_p$ is frozen, and low-rank adaptation is applied to adapt the singular vectors of $\bm{W}_p$ during fine-tuning. SpectralFT aims to effectively capture task-specific knowledge during fine-tuning while preserving and leveraging the pre-trained information.
     
\section{Methodology}
\begin{figure*}[t]
\centering
\includegraphics[width= 0.96 \textwidth]{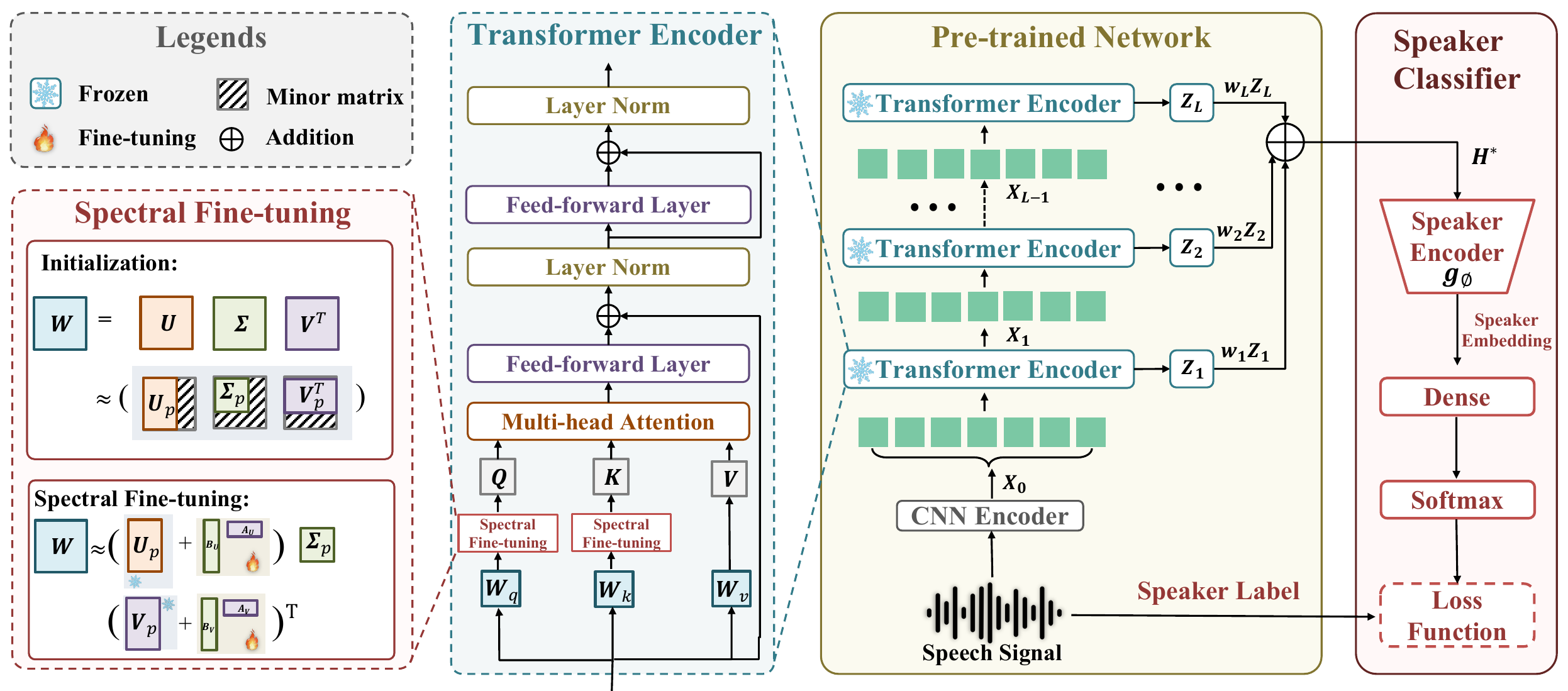}
\caption{The architecture of the proposed SpectralFT. The principal singular components $(\bm{U}_p, \bm{V}_p, \bm{\Sigma}_p)$ are retained to form a low-rank approximation of the original weight matrix $\bm{W}$, which is then fine-tuned using the principle of LoRA. During fine-tuning, only the low-rank matrices $\bm{B}_U$, $\bm{A}_U$, $\bm{B}_V$, and $\bm{A}_V$ are updated, while the principal matrices $\bm{U}_p$ and $\bm{V}_p$ remain frozen. For the operations and principles of the Transformer Encoder, Pre-trained Network, and Speaker Classifier, readers are referred to \cite{li2024dual,li2024parameter}.}
\label{fig:spectralft}
\end{figure*}

As shown in Fig.~\ref{fig:spectralft}, we utilize SVD to decompose the pre-trained weight matrices, exploring the mechanisms of LoRA within the SVD framework. Our method strikes a good balance between preserving the generalization capacity of the pre-trained parameters and enabling task-specific adaptation. 

\subsection{Low-Rank Adaptation}
LoRA \cite{hulora} assumes that the updates to a pre-trained weight matrix $\bm{W}_0 \in \mathbb{R}^{m \times n} $ are low-rank, thereby allowing the changes to be represented by two trainable low-rank matrices: $\bm{B} \in \mathbb{R}^{m \times r} $ and $\bm{A} \in \mathbb{R}^{r \times n}$. Specifically, the updated weight matrix is expressed as: 
\begin{equation}
\bm{W} = \bm{W}_{0} + \Delta \bm{W} = \bm{W}_{0} + \frac{\alpha}{r} \bm{BA},
\label{eq:lora}
\end{equation}
where $ \Delta \bm{W} $ represents the weight updates. Here, $ \alpha $ and $ r $ are hyperparameters controlling the scale and the LoRA rank, respectively, with $ r \ll \min(m, n) $.

The pre-trained matrix $ \bm{W}_{0} $ remains fixed during fine-tuning, which significantly reduces the number of trainable parameters, as both $\bm{A} $ and $\bm{B} $ are low-rank matrices. The $\bm{B} $ matrix is initialized to zero, while the $\bm{A} $ matrix is initialized using a Gaussian distribution with zero mean and unit variance. This initialization strategy ensures that $\Delta \bm{W} = \bm{0} $ at the start of fine-tuning. Because LoRA only modifies the linear matrices in the Transformer model, the low-rank matrices $\bm{BA}$'s can be seamlessly merged into the pre-trained linear matrices. This property results in no additional computation or GPU memory during inferencing.

However, the vanilla LoRA method, which constrains updates to a fixed low-rank subspace, presents a significant limitation. Specifically, the low-rank nature of LoRA restricts the difference between the fine-tuned weight matrix $\bm{W}_{0} + \frac{\alpha}{r} \bm{BA} $ and the pre-trained weights $\bm{W}_{0} $ to a low-rank matrix. This constraint severely limits LoRA's ability to fine-tune a model to arbitrary target tasks.

\subsection{Singular Value Decomposition}
Given a matrix $\bm{W} \in \mathbb{R}^{m \times n} $, its SVD is denoted as $\bm{W} = \bm{U} \bm{\Sigma} \bm{V}^\textsf{T}$, where $\bm{U} = [\bm{u}_1, \bm{u}_2, \ldots, \bm{u}_m] \in \mathbb{R}^{m \times m} $ and $\bm{V} = [\bm{v}_1, \bm{v}_2, \ldots, \bm{v}_n] \in \mathbb{R}^{n \times n}$. The columns of $\bm{U}$ are the left singular vectors, and the columns of $\bm{V} $ are the right singular vectors. The diagonal matrix $\bm{\Sigma} \in \mathbb{R}^{m \times n} $ contains the singular values of $\bm{W} $ in descending order. 

This decomposition can also be reformulated in matrix form. The matrix $\bm{U}$ can be column-wise partitioned into a {\it p}rincipal matrix and a {\it m}inor matrix: $\bm{U}=[\bm{U}_p,\bm{U}_m]$, where $\bm{U}_p = [\bm{u}_1, \bm{u}_2, \ldots, \bm{u}_k]$ and $\bm{U}_m = [\bm{u}_{k+1}, \bm{u}_{k+2}, \ldots, \bm{u}_m] $ are the left singular vectors corresponding to the principal and minor singular values, respectively.\footnote{The subscript of a matrix (e.g., $p$ and $m$ in $\bm{U}_p$ and $\bm{U}_m$) is used for naming the matrix, whereas the subscript of a vector (e.g., $k$ in $\bm{u}_k$) represents the vector's position in a matrix.} The matrices $\bm{V} $ and $\bm{\Sigma} $ are partitioned similarly. Thus, the SVD of $\bm{W} $ can be expressed as:
\begin{equation}
\bm{W}= \bm{U} \bm{\Sigma} \bm{V}^\textsf{T} = \bm{U}_p \bm{\Sigma}_p \bm{V}_p^\textsf{T} + \bm{U}_m \bm{\Sigma}_m \bm{V}_m^\textsf{T} = \bm{W}_p + \bm{W}_m.
\label{eq:svd}
\end{equation}

\subsection{Spectral Fine-tuning}
Inspired by the parameter efficiency of LoRA and the close connection between matrix rank and spectral representation, we explore a spectral fine-tuning mechanism. The idea is to apply SVD to a pre-trained model's weight matrix, followed by fine-tuning the principal columns of the singular vector matrices. To this end, we approximate the SVD of a weight matrix $\bm{W}$ by the spectral representation of $\bm{W}_p$ in Eq.~\ref{eq:svd}, i.e., $\bm{W} = \bm{U} \bm{\Sigma} \bm{V}^\textsf{T} \approx \bm{U}_p \bm{\Sigma}_p \bm{V}_p^\textsf{T}$. We define the additive spectral adapter as
\begin{equation}
\begin{aligned}
\text{SpectralFT}(\bm{W}) :&= [\bm{U}_p + \bm{\Delta}_U] \bm{\Sigma}_p [\bm{V}_p + \bm{\Delta}_V]^\textsf{T},
\end{aligned}
\label{eq:SpectralFT}
\end{equation}
where $\bm{U}_p \in \mathbb{R}^{m \times k}$ and $\bm{V}_p \in \mathbb{R}^{n \times k}$ represent the top-$k$ columns of $\bm{U} $ and $\bm{V} $, respectively. The adaptation set $\bm{\Delta} =\{\bm{\Delta}_U,\bm{\Delta}_V\}$ consists of trainable matrices with the same dimensions as $\bm{U}_p$ and $\bm{V}_p$, respectively. As observed in LASER \cite{sharmatruth}, the minor singular components of a weight matrix often contain noisy information, whereas the principal singular components capture important features across tasks. Therefore, we discard $\bm{U}_m$ and $\bm{V}_m$ in Eq.~\ref{eq:svd}. 

To leverage the advantage of LoRA, we define $\bm{\Delta}_U \equiv \frac{\alpha}{r}\bm{B}_U\bm{A}_U$, where $\bm{B}_U \in \mathbb{R}^{m \times r}$ and $\bm{A}_U \in \mathbb{R}^{r \times k}$, such that $r \ll k $. The matrix $\bm{B}_U$ is initialized to zero, while $\bm{A}_U$ is initialized using a Gaussian distribution. The adapter weights $\bm{B}_U$ and $\bm{A}_U$ are initialized such that $\bm{B}_U \bm{A}_U = \bm{0}$. The same strategy is applied to $\bm{\Delta}_V \equiv \frac{\alpha}{r} \bm{B}_V \bm{A}_V$, where $\bm{B}_V \in \mathbb{R}^{n \times r}$ and $\bm{A}_V \in \mathbb{R}^{r \times k}$. During training, only the elements of $\bm{B}_U$, $\bm{A}_U$, $\bm{B}_V$, and $\bm{A}_V$ are updated.

\subsection{Computation Considerations}
We propose incorporating spectral information into the fine-tuning process for the $\bm{W}_q $ and $\bm{W}_k $ matrices in the attention mechanism of the Transformer model. Our method allows for flexible parameter budgets by adjusting the values of $r$ and $k$. Specifically, we fine-tune the top-$k$ columns of $\bm{U}$ and $\bm{V}$ using additive tuning, which requires storing only $\bm{B}_U$, $\bm{A}_U$, $\bm{B}_V$, and $\bm{A}_V$. 

The only overhead is the runtime and GPU storage during training. Because our method involves only matrix multiplication during the forward pass, it should run as efficiently as LoRA. While the SVD process may introduce some runtime overhead, it is a one-time operation per model and can be reused for subsequent fine-tuning on different downstream tasks.

\section{Experiments and Results}
\subsection{Implementation Details}
We selected HuBERT-Large \cite{hsu2021hubert} and WavLM-Large \cite{chen2022wavlm} as the pre-trained models (PTMs) and ECAPA-TDNN \cite{desplanques20_interspeech} as the speaker encoder. VoxCeleb1-dev \cite{nagrani2017VoxCeleb} and CN-Celeb1 \cite{fan2020cn} were used to fine-tune the PTMs and train the ECAPA-TDNN. We truncated each training utterance to 2 seconds and used mini-batches of 256 utterances for fine-tuning and training. AAM-Softmax \cite{deng2019arcface} was employed, with a margin of 0.2 and a scaling factor of 30. The rank $r$ was set to 16, and the number of top singular vectors $k$ was 256.

\subsection{Results and Analysis}
Table~\ref{tab:result} shows that utilizing a pre-trained model for frame-level feature extraction enhances SV performance (compare Rows 1, 2, and 3), especially after fine-tuning the pre-trained models. We compare our approach with three widely used parameter-efficient fine-tuning methods: Adapter \cite{houlsby2019parameter} (results extracted from \cite{li2024parameter}), static prompt tuning \cite{li2024dual} (results extracted from \cite{li2024parameter}), and LoRA (results extracted from \cite{li2024parameter}) which was used to fine-tune the $\bm{W}_q$, $\bm{W}_k$, and $\bm{W}_v$ matrices in the attention mechanism, with the scaling factor ($\frac{\alpha}{r}$ in Eq.~\ref{eq:lora}) set to 0.1. The results demonstrate that our proposed method outperforms all others on both datasets, with the improvement being particularly pronounced compared to traditional LoRA. This advantage arises from the SVD being able to preserve the most critical features relevant to speaker characteristics while ignoring the unimportant factors that may negatively affect speaker verification. Therefore, the SVD provides a top spectral space that is more relevant to speakers for LoRA-style fine-tuning. With $k\gg r$, SpectralFT can maintain sufficient spectral contents without overparameterizing the LoRA adaptation matrix, an important advantage of SpectralFT over conventional LoRA.

\begin{table*}[t]
    \centering
    \caption{Performance on the test sets of VoxCeleb1 and CN-Celeb1, using HuBERT-Large or WavLM-Large as PTM and ECAPA-TDNN as the speaker encoder. Row 1 uses Filterbank features as input to the ECAPA-TDNN. Results based on full fine-tuning are in italics. They are expected to be the best. The best results based on other fine-tuning methods are in bold.}
    \begin{tabular}{|c|c|c|c|c|c|c|}
        \hline 
        \multirow{2}{*}{PTM} & \multirow{2}{*}{Row} & \multirow{2}{*}{Fine-tuning Method} & \multicolumn{2}{c|}{VoxCeleb1-O} & \multicolumn{2}{c|}{CN-Celeb1} \\
        \cline{4-7} 
        & & & EER(\%) & minDCF &  EER(\%) & minDCF  \\
        \hline
        None & 1 & None & 2.96 & 0.30 & 12.49 & 0.67 \\
        \hline 
        \multirow{6}{*}{HuBERT-Large} & 2 & None & 2.76 & 0.30 & 12.05 & 0.61 \\
        & 3 & Full fine-tuning  & \textit{1.98} &\textit{0.22} & \textit{10.51} & \textit{0.60} \\
        & 4 & Adapter \cite{li2024parameter}  & \textbf{2.13} & 0.24 & 10.89 & 0.62 \\
        & 5 & Static prompt tuning \cite{li2024parameter} & 2.26& 0.23 & 10.69 & 0.59 \\
        & 6 & LoRA ($r$=16, $\frac{\alpha}{r}$=0.1) \cite{li2024parameter} & 2.38 & 0.23 & 10.48 & 0.60 \\
        & 7 & \textbf{SpectralFT (Ours)} & 2.31 & \textbf{0.22} & \textbf{10.45} & \textbf{0.58} \\
        \hline
        \multirow{6}{*}{WavLM-Large} & 8&  None & 1.94 & 0.22 & 11.17 & 0.59 \\
        & 9 & Full fine-tuning  &  \textit{1.39} & \textit{0.16} & \textit{10.47} & \textit{0.56} \\
        & 10 & Adapter \cite{li2024parameter} &  1.68 & 0.19 & 10.83 & 0.63 \\
        & 11 & Static prompt tuning \cite{li2024parameter} &  1.65& 0.18 & 10.57 & 0.58 \\
        & 12 &  LoRA ($r$=16, $\frac{\alpha}{r}$=0.1) \cite{li2024parameter} & 1.88 & 0.21 & 10.89 & 0.63 \\
        & 13 & \textbf{SpectralFT (Ours)} & \textbf{1.47} & \textbf{0.16} & \textbf{10.69} & \textbf{0.56} \\
        \hline
    \end{tabular}
    \label{tab:result}
\end{table*}

\subsection{Investigating Different Rank Settings}
\begin{figure}[t]
    \centering
    \begin{minipage}[b]{0.45\linewidth}
        \centering
        \includegraphics[width=\linewidth]{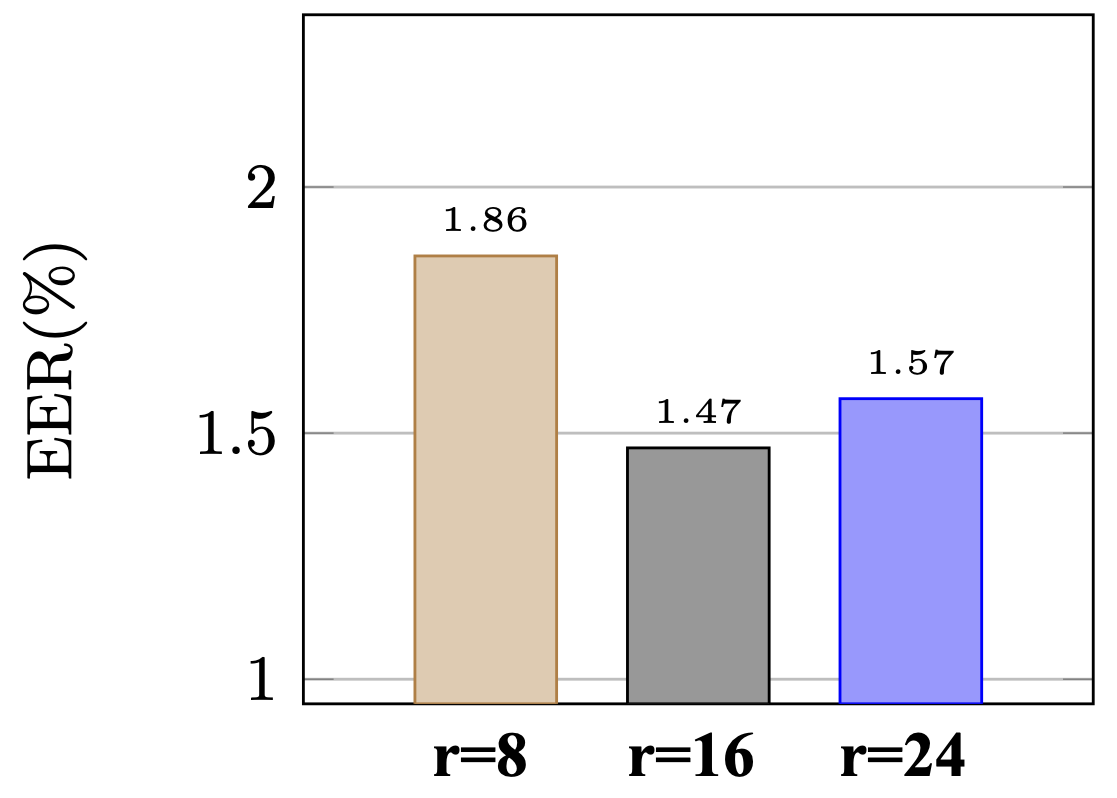}
    \end{minipage}
    \begin{minipage}[b]{0.45\linewidth}
        \centering
        \includegraphics[width=\linewidth]{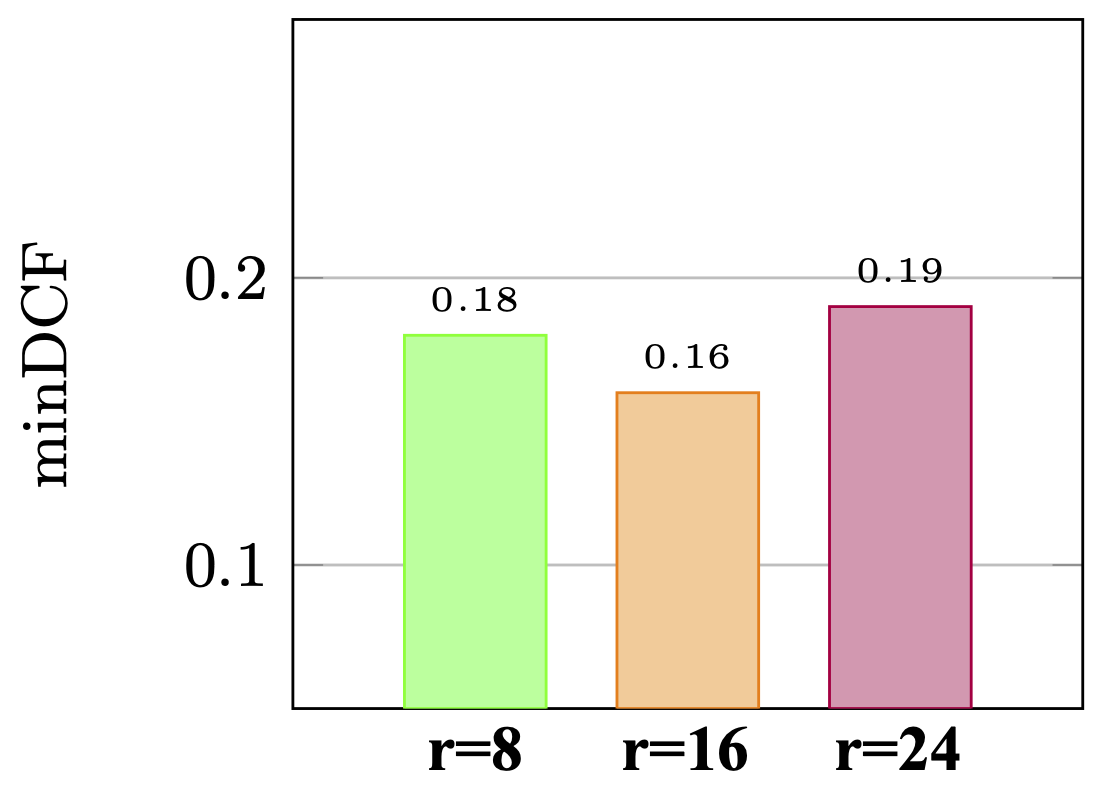}
    \end{minipage}
    \caption{Results on VoxCeleb1-O for different ranks, using WavLM-Large as the PTM.}
    \label{fig:rank}
\end{figure}

We examined the impact of varying the rank $r$ on the fine-tuned WavLM-Large model. As shown in Fig.~\ref{fig:rank}, SpectralFT with a rank of 16 yielded the best performance. The results indicate that selecting an appropriate rank is crucial for good performance when fine-tuning with SpectralFT. Insufficient rank means the subspace for fine-tuning the weight matrices is too restrictive, causing the fine-tuned model to fail to adapt to the downstream task. Conversely, while a higher rank allows the model to capture more details about the downstream task, it may also result in overfitting by learning noise from the adaptation data. Our results show that a rank of 16 strikes a good balance, suggesting that a moderate model capacity is sufficient to capture key features while maintaining strong generalization ability.

\subsection{Analysis of Principle Columns}
\label{sec:principle}
We conduct experiments to investigate the influence of the number of singular components on fine-tuning performance. We set the dimensions of the retained primary singular value components ($k$) to 64, 128, 256, 512, and 1024. Table~\ref{tab:principle} shows that the best results are achieved when retaining 256 components. A spectral space with 256 dimensions is enough because beyond which the singular values are too small for the spectral space to focus on the speaker features. The variation in the low spectral space contains more noise, which could interfere with the speaker verification task.
\begin{table}[t]
    \centering
    \caption{Results on VoxCeleb1-eval using different number of principal columns ($k$) in $\bm{U}$.}
    \begin{tabular}{c|c|c}
        \hline 
        \multirow{2}{*}{No. of Principal columns $k$} & \multicolumn{2}{c}{VoxCeleb1-O} \\
        \cline{2-3}
        & EER(\%) & minDCF \\
        \hline
        64   & 1.83 & 0.22 \\
        128  & 1.59 & 0.21 \\
        256  & 1.47 & 0.16 \\
        512  & 1.51 & 0.16 \\
        1024 & 1.58 & 0.18 \\
        \hline
    \end{tabular}
    \label{tab:principle}
\end{table}

\subsection{Analysis of the Effect of Singular Vectors}
To explore the effect of different singular value settings, we conducted experiments in which only the principal singular components were retained, and we denoted the subspace as ``$\bm{U}_p \bm{\Sigma}_p \bm{V}_p^\textsf{T}$''. We explored the effect of having $\bm{\Delta}_U$ and $\bm{\Delta}_V$ in Eq.~\ref{eq:SpectralFT}. In the third row of Table~\ref{tab:minor}, we used both the principal and minor singular components, fine-tuning the primary singular value components $\bm{U}_p$ and $\bm{V}_p$, while keeping the minor components $\bm{U}_m$ and $\bm{V}_m$ frozen. In the fourth row of Table~\ref{tab:minor}, we considered performing SVD on the weight matrices as the baseline and denoted it as ``$\bm{U} \bm{\Sigma} \bm{V}^\textsf{T}$''. 

The results presented in Table~\ref{tab:minor}, comparing the first and second rows, illustrate the effectiveness of applying our SpectralFT method. Comparing the first and third rows indicates that incorporating $\bm{U}_m$ and $\bm{V}_m$ led to a decline in performance, as $\bm{U}_m$ and $\bm{V}_m$ introduced more speaker verification-unfavorable noise. Comparing the first and fourth rows demonstrates that retaining the principal singular components, discarding minor singular components, and applying SpectralFT can significantly improve performance.

\begin{table}[t]
    \centering
    \caption{Results of different subspace fine-tuning strategies on VoxCeleb1-eval, using WavLM-Large as the PTM.}
    \resizebox{\columnwidth}{!}{%
    \begin{tabular}{c|c|c|c|c}
        \hline 
        \multicolumn{2}{c|}{Subspace} & \multirow{2}{*}{\begin{tabular}{@{}c@{}}$\bm{\Delta}_U$ and $\bm{\Delta}_V$ \\ (in Principal Subspace)\end{tabular}} & \multicolumn{2}{c}{VoxCeleb1-O} \\
        \cline{1-2} \cline{4-5}
        Principal & Minor & & EER(\%) & minDCF \\
        \hline
        $\bm{U}_p \bm{\Sigma}_p \bm{V}_p^\textsf{T}$ &None &\ding{51} & 1.47 & 0.16 \\
        $\bm{U}_p \bm{\Sigma}_p \bm{V}_p^\textsf{T}$ & None &\ding{55} & 1.60 & 0.17 \\
        $\bm{U}_p \bm{\Sigma}_p \bm{V}_p^\textsf{T}$ & $\bm{U}_m \bm{\Sigma}_m \bm{V}_m^\textsf{T}$&\ding{51} & 1.65 & 0.17 \\
        \cline{1-3}
        \multicolumn{2}{c|}{$\bm{U} \bm{\Sigma} \bm{V}^\textsf{T}$} &\ding{55} & 1.68 & 0.20 \\
        \hline
    \end{tabular}
    }
    \label{tab:minor}
\end{table}

\subsection{Analyze the Fine-tuning Positions}
To identify the most effective weight matrices for spectral fine-tuning, we apply SpectralFT progressively to $\bm{W}_q$, $\bm{W}_k$, and $\bm{W}_v$ in the Transformer attention mechanism. We also compared the results with other low-rank approximation fine-tuning methods, specifically LoRA and DoRA. In Table~\ref{tab:position}, $r$ represents the rank, and $\alpha$ represents different scaling factors in Eq.~\ref{eq:lora}. The experimental results indicate that the best performance is achieved when fine-tuning the $\bm{W}_q$ and $\bm{W}_k$ matrices. In Transformer-based models, the $\bm{W}_q $ and $\bm{W}_k $ matrices are responsible for computing attention scores, which determine how the model selects information from the input data. By adjusting the $\bm{W}_q$ and $\bm{W}_k $ matrices, SpectralFT can more precisely control the attention without altering the value matrix $\bm{W}_v$.
\begin{table}[t]
    \centering
    \caption{Results on the test sets of VoxCeleb1 with fine-tuning different weight matrices.}
    \begin{tabular}{c|c|c|c|c|c}
        \hline 
        \multirow{2}{*}{Methods}& \multicolumn{3}{c|}{Weight Type} & \multicolumn{2}{c}{VoxCeleb1-O} \\
        \cline{2-6}
        & $\bm{W}_q$& $\bm{W}_k$& $\bm{W}_v$& EER(\%) & minDCF \\
        \hline
        \multirow{3}{*}{LoRA ($r$=16, $\frac{\alpha}{r}$=1)} & \ding{51} & \ding{55} & \ding{55} & 1.59 & 0.19 \\
        & \ding{51} & \ding{51} & \ding{55} & 1.58 & 0.18 \\
        & \ding{51} & \ding{51} & \ding{51} & 1.88 & 0.21 \\
        \hline
        \multirow{3}{*}{DoRA ($r$=16)} & \ding{51} & \ding{55} & \ding{55} & 1.67 & 0.19 \\
        & \ding{51} & \ding{51} & \ding{55} & 1.54 & 0.17 \\
        & \ding{51} & \ding{51} & \ding{51} & 1.65 & 0.18 \\
        \hline
        \multirow{3}{*}{SpectralFT ($r$=16, $\frac{\alpha}{r}$=1)} & \ding{51} & \ding{55} & \ding{55} & 1.60 & 0.18 \\
        & \ding{51} & \ding{51} & \ding{55} & 1.47 & 0.16 \\
        & \ding{51} & \ding{51} & \ding{51} & 1.64 & 0.19 \\
        \hline
    \end{tabular}
    \label{tab:position}
\end{table}

\section{Conclusions}
In this work, we explore integrating spectral information from the pre-trained model weight matrices into existing PEFT by introducing a spectral adaptation mechanism that updates only the top singular vectors of the pre-trained weight matrices. Empirically, we demonstrate the superiority of our proposed spectral adaptation method over various recent PEFT approaches through extensive experiments.

\bibliographystyle{IEEEtran}
\bibliography{ref}

\end{document}